\documentclass[12pt, prd, showpacs]{revtex4}
\usepackage{amssymb}
\usepackage{amsmath}

\input{tcilatex}

\begin{document}

\title{Negative energy states in the Reissner-Nordstr\"{o}m metric}
\author{O. B. Zaslavskii}
\affiliation{Department of Physics and Technology, Kharkov V.N. Karazin National
University, 4 Svoboda Square, Kharkov 61022, Ukraine}
\affiliation{Institute of Mathematics and Mechanics, Kazan Federal University, 18
Kremlyovskaya St., Kazan 420008, Russia}
\email{zaslav@ukr.net }

\begin{abstract}
We consider electrogeodesics on which the energy $E<0$ in the
Reissner-Nordstr\"{o}m metric. It is shown that outside the horizon there is
exactly one turning point inside the ergoregion for such particles. This
entails that such a particle passes through an infinite chain of black-white
hole regions or terminates in the singularity. These properties are relevant
for two scenarios of high energy collisions in which the presence of white
holes is essential.
\end{abstract}

\keywords{particle collision, super-Penrose process, white holes}
\pacs{04.70.Bw, 97.60.Lf }
\maketitle

\section{Introduction}

One of the most physically interesting processes in black hole physics is
the famous Penrose process \cite{pen}. It consists in the possibility of
energy extraction from black holes. This requires the existence of the
negative energy states with respect to a remote observer. Then, if some
particle having the energy $E_{0}$ decays to two fragments and particle 1
sits on the orbit with the energy $E_{1}<0$, particle 2 can return to
infinity with $E_{2}>E_{0}$. The region where such states can exist is
called ergoregion (ergosphere). It can be realized in the metric of rotating
black holes - the simplest example is the Kerr one. For the Schwarzschild
metric such an effect is absent. Meanwhile, it was found that the
counterpart of the Penrose process for static but electrically charged black
holes described by the Reissner-Nordstr\"{o}m (RN)\ metric, is also \
possible \cite{ruf}. Many aspects of the Penrose process were investigated
but, strange as it may seem, the question about the nature of trajectories
with $E<0$, remained in the shade until recently. It was posed in \cite%
{gpneg} where it was shown that corresponding geodesics must originate and
terminate under the horizon. This means that the situation when in the
ergoregion a particle with $E<0$ oscillates between two turning points
outside the horizon or moves along the circular orbit, is forbidden. This
result was extended to a rather wide class of stationary rotating axially
symmetric black holes in \cite{neg15}.

The aim of the present work is to elucidate this issue for the RN metric.
The trajectories under discussion play in the electric version of the
Penrose process \cite{ruf} a role similar to that played by corresponding
geodesics in the standard Penrose process \cite{pen}. An additional
motivation stems from the fact that (i) the collisional version of the
Penrose process also exists (see \cite{col} for a review and (ii) it can be
related to the high energy processes in the centre of mass frame in the
vicinity of rotating black holes \cite{ban}. The latter process is called
the Ba\~{n}ados-Silk-West (BSW) effect under the names of its authors. The
counterpart of the BSW effect as well as a version of the collisional
Penrose process (even with formally unbounded efficiency) are possible not
only for rotating black holes but also for the RN metric \cite{jl}, \cite{rn}%
. Moreover, recently the analogue of the latter was considered even for the
white charged holes in the extremal case \cite{whq} (that has also its
counterpart in the Kerr case \cite{ph}). Now, we consider a generic
(nonextremal or extremal) RN black hole.

We use the system if units in which fundamental constants $G=c=1$.

\section{Basic equaitons}

Let us consider motion of particle with the mass $m$ and electirc charge $q$
in the backgrounf of the RN black hole. The metric has the form%
\begin{equation}
ds^{2}=-fdt^{2}+\frac{dr^{2}}{f}+r^{2}d\omega ^{2}\text{, }d\omega
^{2}=d\theta ^{2}+\sin ^{2}\theta d\phi ^{2}\text{,}
\end{equation}%
\begin{equation}
f=1-\frac{2M}{r}+\frac{Q^{2}}{r^{2}}\text{,}
\end{equation}%
where $M$ is the black hole mass, $Q$ being its electric charge. (For the
definiteness we assume that $Q>0$.) The event horizon $r_{+}$ lies at the
largest rot of the equation $f=0$,%
\begin{equation}
r_{+}=M+\sqrt{M^{2}-Q^{2}}\text{.}  \label{hor}
\end{equation}

Then, the equations of motion read%
\begin{equation}
m\dot{t}=\frac{X}{f}\text{,}
\end{equation}%
\begin{equation}
X=E-q\varphi \text{,}
\end{equation}%
\begin{equation}
m\dot{r}=\sigma P\text{, }\sigma =\pm 1\text{,}
\end{equation}%
\begin{equation}
P=\sqrt{X^{2}-f(m^{2}+\frac{L^{2}}{r^{2}})}\text{.}  \label{P}
\end{equation}%
Here, $E$ is the energy$,L$ being the angular momentum. They are conserved
since the metric does not depend on $t$ and $\phi $. Dot denotes derivative
with respect to the proper time $\tau $. The electric Coloumb potential $%
\varphi =\frac{Q}{r}$.

The forward-in-time condition $\dot{t}>0$ gives us%
\begin{equation}
X\geq 0\text{,}  \label{ft}
\end{equation}%
where equality can be reached on the horizon only, where $f=0$.

\section{Proof of the statement}

We are interested in the states with nonpositive energies, so%
\begin{equation}
E=-\left\vert E\right\vert \text{.}
\end{equation}%
Then, (\ref{ft}) gives us $q=-\left\vert q\right\vert <0$, so%
\begin{equation}
X=\frac{Q\left\vert q\right\vert }{r}-\left\vert E\right\vert \text{.}
\end{equation}%
A particle with $E<0$ cannot reach infinity since it would have there $X<0$
in contradiciton with (\ref{ft}). It means that in the region $r>r_{+}$ it
has a turning point $r_{1}$, where $P(r_{1})=0$. Now, we will show that in
this region it has only one such a point. To this end, we will demonstrate
that $\left( P^{2}\right) ^{\prime }<0,$ prime denotes $\frac{d}{dr}$. Then,
the function $P(r)$ is monotonically decreasing in this region, so $%
P(r_{+})\geq P(r)\geq P(r_{1})=0.$

The proof is rather easy, despite some algebra. Direct calculation gives us%
\begin{equation}
\frac{\left( P^{2}\right) ^{\prime }}{2}=\frac{BL^{2}}{r^{3}}-\frac{%
m^{2}(Mr-Q^{2})}{r^{3}}-\frac{Q\left\vert q\right\vert }{r^{2}}X\text{,}
\label{pp}
\end{equation}%
where%
\begin{equation}
B=f+\frac{Q^{2}}{r^{2}}-\frac{M}{r}=1-\frac{3M}{r}+\frac{2Q^{2}}{r^{2}}.
\end{equation}%
Here, $Mr-Q^{2}\geq Mr_{+}-Q^{2}\geq 0$. As a result,%
\begin{equation}
B\leq f.  \label{bf}
\end{equation}

If $B\leq 0$, the fact that $\left( P^{2}\right) ^{\prime }<0$ is obvious
since all terms in (\ref{pp}) are negative outside the horizon. We assume
that $B>0$ and will continue derivation.

It follows from $P^{2}\geq 0$ that%
\begin{equation}
\frac{L^{2}}{r^{2}}\leq \frac{\left( \frac{Q\left\vert q\right\vert }{r}%
-\left\vert E\right\vert \right) ^{2}-fm^{2}}{f}.
\end{equation}

Then, by substitution into (\ref{pp}) we obtain that 
\begin{equation}
\frac{\left( P^{2}\right) ^{\prime }}{2}\leq \frac{C}{rf}\text{,}
\end{equation}%
where%
\begin{equation}
C=B\left( \frac{Q\left\vert q\right\vert }{r}-\left\vert E\right\vert
\right) ^{2}-fD\text{,}
\end{equation}%
\begin{equation}
D=m^{2}f+\frac{Q\left\vert q\right\vert }{r}(\frac{Q\left\vert q\right\vert 
}{r}-\left\vert E\right\vert )\text{.}
\end{equation}%
Taking into aacount (\ref{bf}), we can rewrite this as%
\begin{equation}
C\leq Hf\text{,}
\end{equation}%
where%
\begin{equation}
H=X^{2}-X\frac{Q\left\vert q\right\vert }{r}-m^{2}=-X\left\vert E\right\vert
-m^{2}<0\text{.}
\end{equation}%
Thus $C<0,$ so $P^{\prime }<0$ that is just what we wanted to prove. Thus,
no more than 1 turning point can exist outside the event horizon for
particles with a negative (or zero) energy. 

From another hand, one turning point in the outer region is inevitable, as
is explained above. It means that a particle with $E<0$ appears in the outer
region from the inner region only for a finite interval of the proper time,
bounces back and moves inside. There, the particle under discussion can
bounce from the singularity and appear in the outer space again but in
another universe \cite{gb}, \cite{bc}. The process can continue endlessly.
Alternatively, such a particle can fall in the singularity (or originate
from it), provided $L=0$ and either (i) $\left\vert q\right\vert >m$ or (ii) 
$\left\vert q\right\vert =m$, $Mm\geq Q\left\vert E\right\vert $. (There is
an exceptional case $\left\vert E\right\vert =m=\left\vert q\right\vert ,M=Q$
when a particle remains in the rest in the field of the extremal black hole.)

\section{Discussion and conclusions}

Thus we showed that for the RN metric the situaiton with the electogeodesics
(trajectories corresponding to motion under the electromagnetic force and
gravitation only) is similar to that in the Kerr metric \cite{gpneg} or more
general rotating black-white holes \cite{neg15}. In the ergosphere, there is
exactly one turning point outside the horizon. It follows from this that a
particle with $E<0$ emerges from the inner (white hole) region and,
afterwards, returns to the inside region under the horizon. It passes
through an infinite chain of different black-white hole region or ends up
hitting the singularity.

The relevance of a white hole region entails, in particular, that \ two
types of high energy collisions connected with white hole are possible. In
the first type of a scenario it can appear from the white hole region and
collide with another particle having $E>0$. If collision occurs near the
horizon, the energy in the center of mass frame becomes unbounded \cite{tot}%
. In the second type of scenario two particle with $E>0$ collide in the
black hole region. They can produce a particle with $E<0$ that oscillates in
the bounded region but passes through different black-white hole regions 
\cite{whq}.

In our treatment, we assumed that the space-time represents the eternal
black-white \ hole. In the case of collapse of charged matter, the
properties of the trajectories under discussion can change. This requires
further investigation.

\begin{acknowledgments}
This work was supported by the Russian Government Program of Competitive
Growth of Kazan Federal University.
\end{acknowledgments}

\end{document}